# Optimal QoS Constraint Service Composition in Mobile Ad Hoc Networks

P Veeresh
*Department of CSE*
*St Johns College of Engineering & Technology*
*Andhra Pradesh, India*
veeresh_kaly@yahoo.co.in

R Praveen Sam
*Department of CSE*
*G. Pulla Reddy Engineering College*
*Andhra Pradesh, India*
praveen_sam75@yahoo.com

Mohammad Riyaz Belgaum
*College of Computer Studies*
*AMA International University*
*Bahrain.*
bmdriyaz@amaiu.edu.bh

*Abstract* – In recent year's computational capability of the mobile nodes have been greatly improved. The mobile nodes have the capability of running different applications. Implementation of services in Mobile Ad Hoc Networks (MANETs) increases the flexibility of using mobile devices for running a wide variety of applications. Single service cannot satisfy the user needs. The complex needs of the users can be satisfied by the service composition. Service composition means, combining the atomic services into a complex service. In this paper we propose QoS constraint service composition in MANETs. We considered both service QoS parameters as well node parameters. Response time and throughput as parameters for services and energy and hop count as node parameters. These four QoS parameters are optimized using a mathematical model Hammerstein model to generate a single output. Based on generated output, max valued (optimal) services are considered in service composition path. The simulation results shown that, our proposed method outperforms than the traditional AODV method of service composition.

## I. INTRODUCTION

In recent years, a rapid development occurred in processing capabilities of mobile devices such as smart phones and personal digital assistants (PDA). Growing needs of the users, it is essential to communicate with each other. Mobile ad hoc network is an infrastructure less, temporary, spontaneous and multi-hop communication network with collection of independent nodes [1]. MANET is a self controlled network and there is no centralized administration. Each node in MANET will have processing capability and as well as forwarding capability. A Mobile Ad Hoc Networks (MANET) constructed through the Wi-Fi or Blue-tooth with the mobile devices such as smart phones, PDAs and laptops. The challenges in MANETs are Dynamic topology, limited bandwidth, packet loss, Battery constraints and Security.

Service Oriented Architecture (SOA) has attractive benefits. The services are the implementation of SOA. Services are self-descriptive, self-encapsulated, dynamic discovery, loosely coupled, heterogeneous, machine interaction and dynamic loading components.

Service oriented methodology retains the benefits of component-based development. A lot of service oriented research is presented in [2],[4],[3]. It has an interface portrayed in a machine-process able format (especially Web Service Description Language-WSDL). In SOA, other systems can interact seamlessly with each other using SOAP messages. Implementing services into mobile ad hoc network increases taste of technology and solves complex tasks of the users [5].

Service discovery means identifying required services in vicinity of network. Service discovery in wired networks uses centralized registries i.e., Universal Description Discovery and Integration-UDDI. The approaches defined for wired networks are not suitable for wireless networks. Due to dynamic nature of topology any node can join and leave network any time. There are no Permanent nodes present in the network to maintain centralized registries in mobile ad hoc networks. Suppose, a node in MANET elected and maintained all services information. If that node moves outside of the vicinity the entire services information gets collapsed and that will not known to anyone.

Service composition means combining individual services into a large and complex distributed service to satisfy the needs of the users. Service composition includes combining and planning an arrangement of services. This process can be performed either automatically or manually [6], but requires co-operation from other mobile nodes.

Wired networks are stable networks, but MANETs are not stable networks. In mobile ad hoc networks energy constraints are very important because nodes in the network purely depend on limited energy sources. Service discovery and composition methodologies defined for wired networks not suitable for Mobile Ad hoc network due to the topology of mobile ad hoc networks is time-varying and dynamic in nature [7]. This may enable great challenges in MANETs for service composition.

To counter the problems of existing approaches, the proposed method has the following features.
1. A decentralized approach for maintains of services information.
2. Using Hammerstein mathematical model QoS parameters are optimized and max trusted services are considered in service composition path.

3. Multiple services can be accommodated in a single node.

Service composition success rate will depends on number of nodes involved in composition [17]. The proposed method increases the system performance with minimum number of nodes in the service composition.

The paper organized as follows. Section II specifies related work. Section III describes system model and proposed method. Section IV Simulation results and section V presents Conclusion of our work.

## II. RELATED WORK

In [8], service composition is proposed by constructing hierarchical task graph using smaller components. In graph a node represents logical services and edge represents data flow between corresponding nodes. In [9], a distributed broker-based service composition protocol is proposed for MANET. The composition is accomplished with localized broadcasting and intelligent selective forwarding technique to minimize composition path length.

In [10] proposes agent based self evolving service composition approach. Five stages of composition process are combined together and consider all phases as a single process. It is a decentralized self evolving approach and studies service relation modification and service migration. This approach is suitable for wired network, but not suitable for MANETs because an integrated five phases of services composition process in one node become in-cumbrance in MANET which contains limited computational capability nodes. In MANETs, maintaining of agents is critical task due to dynamic movement of nodes.

In the paper [11], the FPN is extended as a DFPN ("Dynamic Fuzzy Petri Net") where each transition has input and output weights to implement the dynamic nature. Here the information is represented in terms of proposition rules and then represented in the form of FPN. The author proposed concurrent reasoning algorithm to explain the behavior of DFPN.

In the paper [12], MANET is modeled as a FPN, where nodes are like places and wireless links are like transitions. The quality parameters are used to calculate the certainty factor in firing of transition. The trustworthy route is estimated by applying FPN principles in MANETs. The route discovery mechanism is also discussed for multicast routing. The route recovery mechanism is also explained with the help of CRA algorithm.

In [13] Minimum Disruption Index evaluated using dynamic programming for optimal solution for service composition. They used MDSCR ("Minimum Disruption Service Composition and Recovery") algorithm for uncertain node mobility. This algorithm predicts the service link life time by approximating the node location and velocity with one hop look-ahead prediction. But they do not consider the arbitrariness of node mobility. The prediction result may be inaccurate.

In [14] they present QASSA, an efficient service selection algorithm which provides ground for QoS-aware service composition in ubiquitous environments. They formulate global QoS requirements for service selection as a set-based optimization problem.

In [18] proposed a dynamic and distributed constraint satisfaction problem is used to compose services in pervasive systems. Disqualified services are identified and recomposed using heuristic algorithms. When service is not available or leaves from the vicinity, it adapts services without restarting service composition process from each time. Limitation of this approach is Multihop composition and adaptation was not considered.

In [19] proposed a decentralized an agent-based integrated self-evolving service composition approach. This approach systematically takes the integration of five stages of service composition into a single compound stage. It is a decentralized and self evolvable approach. This approach is not well suitable where the computational resource is scarce, especially for MANETs.

## III. SYSTEM MODEL

The proposed method is an Optimal QoS constraint service composition. Here two parameters Response time and Reliability of a service considered as QoS metrics for services and Energy and Node reliability is considered as a QoS metrics for a node.

Service composition initiator issues a complex request and in turn the request is splits into atomic requests. Atomic services requests are transmitted to nearest local service repository nodes. Service repository nodes forward the request to their nearest service repositories. Service repository nodes collect services information in nodes which is in one hop distance. The request is to be forwarded until its TTL ("Time to Live") value reaches to 0 to avoid flooding of packets in the network. If a repository contains required service for composition, it responds with the reply packet. Replay packet contains service related information like node id, service id, service response time, service reliability, node energy, node reliability etc., These information is collected by the composition initiator. All these parameters are considered as input to the Hammerstein model to calculate trust factor of services as shown in figure 1.

After receiving all responses the initiator node runs service composition algorithm to find maximum trusted value services to create service composition path.

### A. Hammerstein model

A Hammerstein model is a Multi Input and Single Output model (MISO)[17]composed of a cascade of two subsystems nonlinear subsystem and linear subsystem. Nonlinear subsystem includes nonlinear gain *functions* $\Psi$ (*) and a dynamic linear part as the linear subsystem. Generally, the system can be modeled by equation (1)

Where $y(t)$ is the system output, and inputs of the system is $u_1, ..., u_m$. $\eta(t)$ is a Gaussian random noise with zero mean and variance of $\sigma^2$. $\Psi(t)$ i = 1, ..., m are the outputs of the nonlinear subsystem and the input to the linear block. $n_a$ and $n_{bi}$ ; i =1, …, m are the input and output lags for the linear subsystem.

## B. Calculation of Node energy

A node can send and receive data packets from the next hop and energy is required to transmit or receive data packet. In paper [15], node energy is calculated as eq(2).

$$E_{total} = 2 \times E_{act} \times k + E_{amp} \times d^2 \times k \qquad (2)$$

Here $E_{act}$ is transmitter/receiver accessing energy. A modulator/ demodulator requires $E_{amp} \times d^2$ amount of energy to transmit K-bit data over d distance.

## C. Evaluation of node reliability

A node can estimate reliability of a node by observing authenticated attitude of forwarding and receiving of packets.

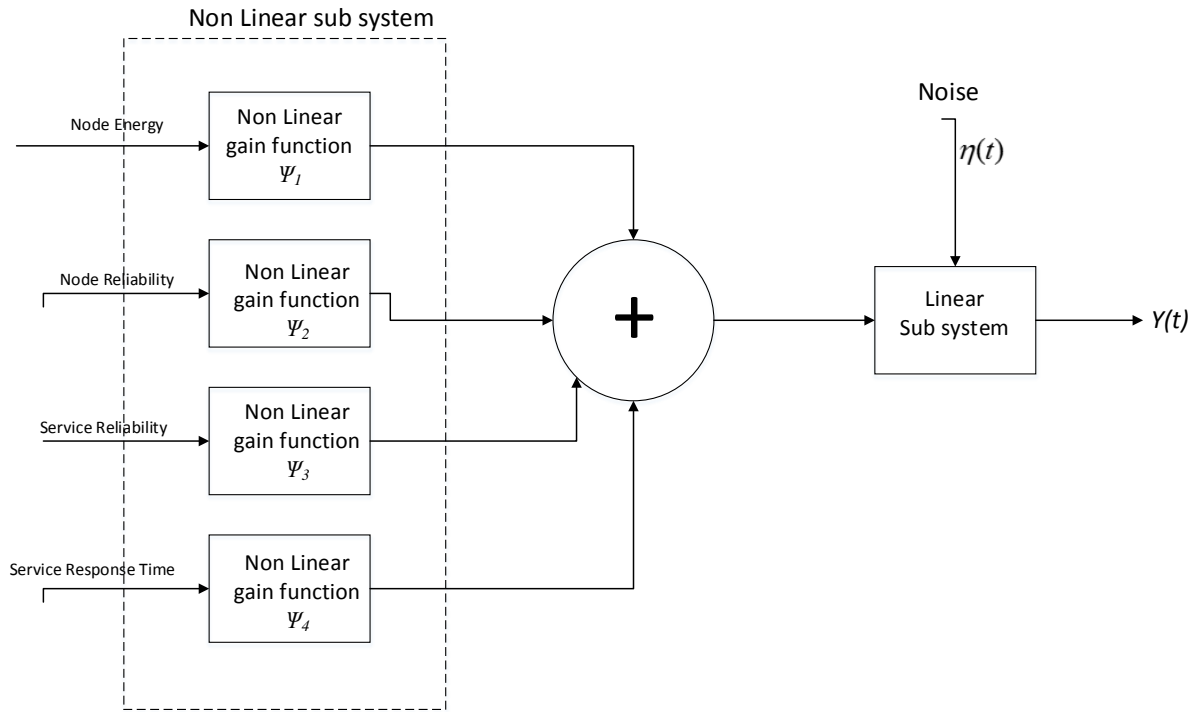

Figure 1 Hammerstein model to calculate trust value of a service

$$y(t) = \sum_{i=1}^{n_a} a_i y(t-i) + \sum_{k=0}^{n_{b_1}} b_k,1\psi 1(u_1(t-k)) + ... + \sum_{k=0}^{n_{b_m}} b_k, m\psi m(u_m(t-k)) + \eta(t) \qquad (1)$$

In paper [16], a node reliability ( r ) is estimated as a random variable using Bayesian inference theory and the value fall between [0, 1]. Lets a node has forwarded $a$ number of packets correctly among the $b$ number of received packets then expectation of reliability is like in eq(3)

$$E[r] = \frac{\alpha_n}{\alpha_n + \beta_n} \qquad (3)$$

Here $\alpha_n = \alpha_{n-1} + a_{n-1}, \beta_n = \beta_{n-1} + b_{n-1} - a_{n-1}$, $\alpha_0 = \beta_0 = 0$

## C. Service Reliability

Web services reliability represents the ability of doing required functions under specified conditions within the given time interval. The reliability is the overall measure of a web service is related to the number of failures per day, week, month, or year. Service reliability evaluated by using equation (4)

$$S(r) = \frac{\#Failures}{Time} \qquad (4)$$

## D. Response Time

Response time is the time interval between sending of request to and receiving of response. Service response time it is divided into service processing time, Network processing time, time consumed for compression and decompression of data, time consumed for encryption and decryption of data and time for data traversing through the protocol stack of source, intermediate and destination nodes.

Response time is represented as:

$$t_{response}(s) = t_{task}(s) + t_{stack}(s) + t_{transport}(s) + t_{cd}(s) + t_{ed}(s) \quad (5)$$

Where

- $t_{response}(s)$ – Response time of a service
- $t_{task}(S)$ – Task Processing Time
- $t_{stack}(S)$ - time consumed for processing of data in protocol stacks of source, intermediate and destination nodes.
- $t_{transport}(t)$ – network transport time
- $t_{cd}(s)$ – Time required for compression and decompression of data.
- $t_{ed}(s)$ – Time required for encryption and decryption.

## E. QoS Constraint Service Composition

In this section, we propose service composition by considering the node constraints as well service constraints. In service discovery process the composition initiator will get necessary information for service composition. Composition initiator will get all Nodes Energy Index value $E_{Total}$ and Node reliability $E[r]$. Composition initiator also gets service response time $t_{resonse}(s)$ & service reliability $S(r)$. These QoS constraints are normalized by using Hammerstein model and a matrix is established between nodes and its services which specifies trust value of service T as shown below.

$$\begin{pmatrix} T_{11} & T_{12} & T_{13} & ... & T_{1n} \\ T_{21} & T_{22} & T_{23} & ... & T_{2n} \\ T_{31} & T_{32} & T_{33} & ... & T_{3n} \\ ... & ... & ... & ... & ... \\ T_{m1} & T_{m2} & T_{m3} & ... & T_{mn} \end{pmatrix}$$

Each row specifies a set of similar services provided by the different nodes with different Trust values. Each column specifies a set of services provided by a node.

The QoS constraint service composition algorithm generates composition path by considering the optimal atomic services provided by the outsourced service providers. First it lists out all the required maximum Hammerstein normalized services $S_1, S_2, S_3, ... S_n$. After that a service composition path and service composition execution plan established. For example composition initiator sends request to service $S_1$ present in node $N_1$. After processing of service $S_1$, $S_1$ handover the composition plan to $S_2$ present in node $N_2$.

For example, matrix consists of trust values of each service.

$$\begin{pmatrix} 58 & 84 & \infty & 48 & 64 \\ 75 & \infty & 80 & 62 & \infty \\ 90 & \infty & \infty & \infty & \infty \\ \infty & 75 & 54 & \infty & \infty \end{pmatrix}$$

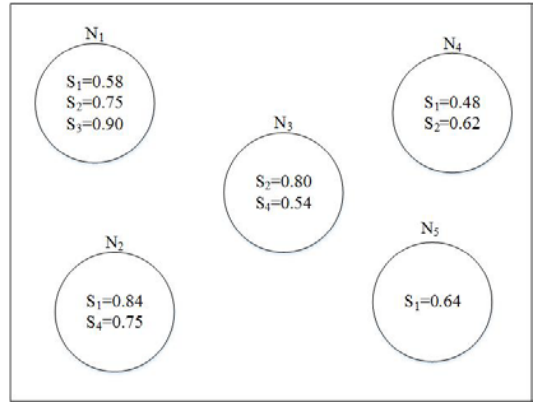

**Figure 2:** Nodes with multiple services and Hammerstein normalized trust values

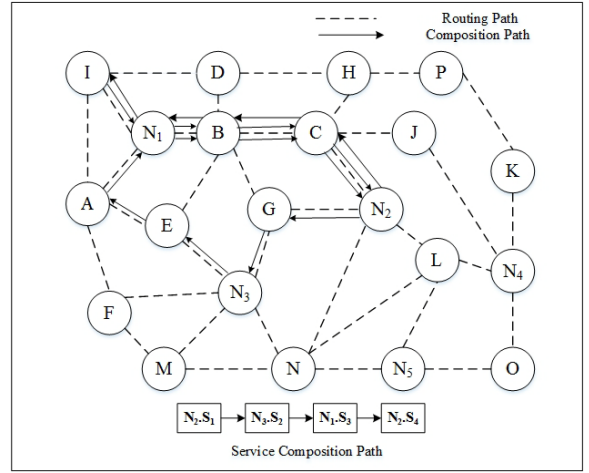

**Figure 3:** Service Composition Path

As shown in Figure 2, Node $N_1$ contains three services $S_1$, $S_2$, $S_3$. Node $N_2$ contains two services $S_1$, $S_4$. Node $N_3$ contains two services $S_2$, $S_4$. Node $N_4$ contains two services $S_1$, $S_2$. Node $N_5$ contains only one service $S_1$. Among these nodes, for Service $S_1$ node $N_2$ selected because maximum Hammerstein normalized trust value service, for Service $S_2$ node $N_3$ selected, for Service $S_3$ node $N_1$ selected, for Service $S_4$ node $N_2$ selected. The optimal service composition path as shown in figure 3 is $N_2 \rightarrow N_3 \rightarrow N_1 \rightarrow N_2$.

## IV. SIMULATION AND PERFORMANCE EVALUATION

In this section, the performance of proposed method is compared with traditional AODV in service composition. The performance is measured with metrics Path Failure rate, Throughput and Service Composition Efficiency. For implementation of services we utilized the tool specified in [13], which is the extension frame work of Network Simulator NS-3. The Table I summarizes Simulation parameters.

TABLE I
Simulation Setup

| Parameter | Value |
|---|---|
| Number of nodes | 100 |
| Simulation Time | 150 Seconds |
| Wifi standard | 802.11b |
| Wifi rate | DsssRate1Mbps |
| Transmission range (R) | 45m |
| Routing protocol | AODV |
| Number of concrete services | 180 |
| Size of composition plan | 5 (Abstract Services) |

TABLE I: Simulation Setup

Firstly, we organized an infrastructure less MANET with 100 mobile devices with wireless capabilities and each device can communicate within their proximity with other devices.

Figure 4 to 6 Represents a) Path Failure rate b) Throughput and c) Service Composition Efficiency. By using Visual Trace Analyzer tool, we obtained the number of times path failed in duration of time as shown in figure 4. The simulation is run for 150 seconds, where the proposed method has shown an enhanced performance when compared with the traditional AODV approach. In the service composition, the proposed method chooses nodes with good energy levels; hence it could reduce the path failures.

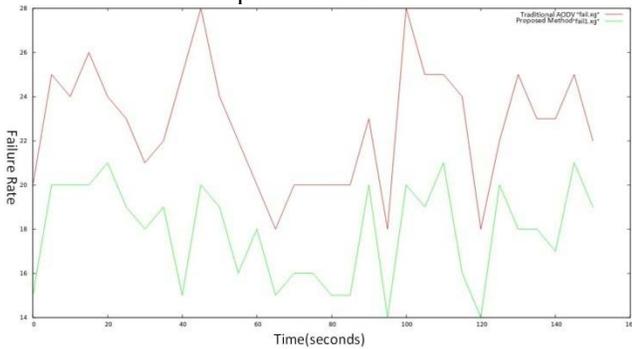

Fig. 4: Service Composition Failure Rate

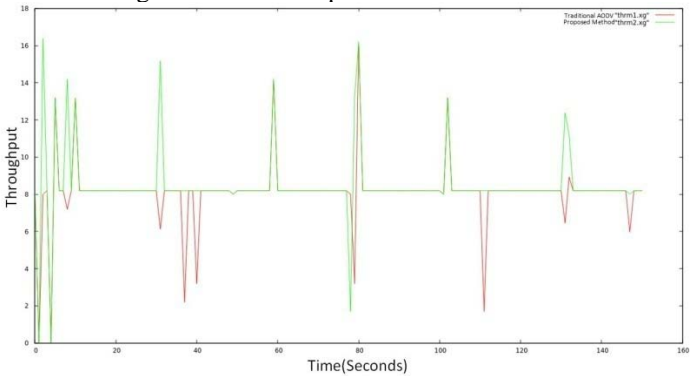

Fig. 5: Throughput

It has shown a rapid increase in the throughput when compared with the traditional AODV approach as shown in figure 5.

In figure 6 Represents proposed approach will selects optimal nodes in all aspects of QoS parameters. It increases the overall composition efficiency then the traditional AODV where it selects without optimization.

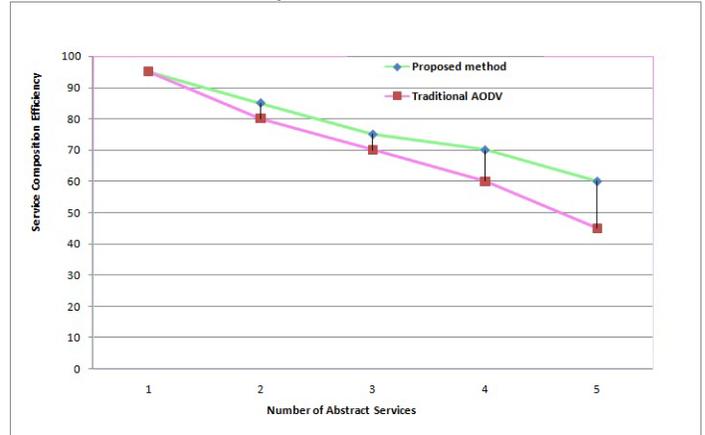

Figure 6. Service Composition Efficiency

V CONCLUSION

The proposed approach Optimal QoS Constraint Services Composition in Mobile Ad Hoc Networks is designed for highly dynamic environment applications, require multiple participants, a node provides multiple services and distribute components in an ad hoc manner. Node mobility, Band width, Energy constraints etc., are the important constraints in MANET. Due course, we proposed Response time and throughput as parameters for services and energy and hop count as node parameters. These QoS parameters are optimized using a mathematical model Hammerstein model to generate a single output. We observed that, our proposed service composition algorithm in MANET performs better results through simulation than traditional AODV based compositions in terms of Service failure rate, Throughput and Service Composition Efficiency.